\documentclass[
showpacs,preprintnumbers,amsmath,amssymb]{revtex4}
\usepackage{amsmath}
\usepackage{graphicx}
\begin{document}

\tolerance=5000

\def\be{\begin{equation}}
\def\ee{\end{equation}}
\def\bea{\begin{eqnarray}}
\def\eea{\end{eqnarray}}
\def\tr{{\rm tr}\, }
\def\nn{\nonumber \\}
\def\e{{\rm e}}

\newcommand{\Eqn}[1]{&\hspace{-0.2em}#1\hspace{-0.2em}&}

\title{Conformal symmetry and accelerating cosmology in teleparallel gravity}
\author{Kazuharu Bamba$^{1,}$\footnote{
E-mail address: bamba@kmi.nagoya-u.ac.jp}, 
Sergei D. Odintsov$^{1, 2, 3, 4,}$\footnote{
E-mail address: odintsov@ieec.uab.es} 
and 
Diego S\'{a}ez-G\'{o}mez$^{1, 5, 6,}$\footnote{
E-mail address: diego.saezgomez@uct.ac.za}
}
\affiliation{
$^1$Kobayashi-Maskawa Institute for the Origin of Particles and the
Universe,
Nagoya University, Nagoya 464-8602, Japan\\ 
$^2$Instituci\`{o} Catalana de Recerca i Estudis Avan\c{c}ats (ICREA), 
Barcelona, Spain\\
$^3$Institut de Ciencies de l'Espai (CSIC-IEEC), 
Campus UAB, Facultat de Ciencies, Torre C5-Par-2a pl, E-08193 Bellaterra
(Barcelona), Spain\\ 
$^4$Tomsk State Pedagogical University, Tomsk, Russia\\
$^5$Astrophysics, Cosmology and Gravity Centre (ACGC) and \\ 
Department of Mathematics and Applied Mathematics, University of Cape Town, Rondebosch 7701, Cape Town, South Africa \\
$^6$Fisika Teorikoaren eta Zientziaren Historia Saila, Zientzia eta Teknologia Fakultatea,\\
Euskal Herriko Unibertsitatea, 644 Posta Kutxatila, 48080 Bilbao, Spain
}

\begin{abstract}
We discuss conformal issues of pure and extended teleparallel gravity. 
In particular, we present formulations of conformal transformation 
in teleparallel gravity. 
Furthermore, we propose conformal scalar and gauge field theories 
in teleparallel gravity and study conformal torsion gravity. 
We explicitly demonstrate that a power-law acceleration 
(including the $\Lambda$CDM universe) as well as
the de Sitter expansion of the universe can be realized 
in extended teleparallel gravity with a conformal scalar field, 
and that pure teleparallel gravity with a conformal scalar field 
can lead to the $\Lambda$CDM model. 
It is also shown that there can exist the de Sitter solution in conformal torsion gravity. 
\end{abstract}
\maketitle 

\section{Introduction}

The cosmological phenomenon of current accelerated expansion of the universe has been revealed by recent cosmological observations including 
Supernovae Ia~\cite{SN1}, large scale structure~\cite{LSS}, 
the baryon acoustic oscillations~\cite{Eisenstein:2005su}, 
cosmic microwave background radiation~\cite{WMAP}, 
and weak lensing~\cite{Jain:2003tba}. 
Provided that the universe is strictly homogeneous, 
there are two representative procedures to study the current cosmic 
acceleration. 
First is to introduce dark energy with 
negative pressure in general relativity 
(for reviews on dark energy, see, e.g.,~\cite{R-DE}). 
Second is to extend or modify gravity on large scales. 

As one of the ways of extending gravitational theories, 
``teleparallelism''~\cite{Teleparallelism, pereira1}, which we call ``teleparallel gravity'' in this paper, has been investigated in various theoretical aspects. 
In teleparallel gravity, 
the torsion scalar $T$, which consists of the Weitzenb\"{o}ck connection, 
describes the action. As a consequence, curvature vanishes, 
although in general relativity (GR), the scalar curvature $R$, which is expressed using the Levi-Civita connection, describes the action. 
It is known that in extended teleparallel gravity, the so-called $f(T)$ gravity, 
whose action is written in terms of a function of $T$, 
not only inflation~\cite{Inflation-F-F} in the early universe but also 
 late time cosmic acceleration~\cite{Bengochea:2008gz, Linder:2010py, WY-BGLL-BGL, Accelerating-cosmology-f(T)-gravity, B-LTCA-f(T)} can be realized. The spirit of this extension of $f(T)$ gravity in teleparallel gravity is the same as that of $F(R)$ gravity in GR 
(for recent review, see Ref.~\cite{F(R)-gravity}). 
An advantage of $f(T)$ gravity is that 
the gravitational field equation is second order as it is 
in GR, while it is forth order in $F(R)$ gravity. 
Hence, it is easier to analyze the cosmological evolution in $f(T)$ gravity, 
in comparison with $F(R)$ gravity, as well as the analysis of cosmological perturbations~\cite{Izumi:2012qj}. 
There are also arguments in terms of theoretical properties of $f(T)$ gravity, 
e.g., local Lorentz invariance~\cite{barrow}, non-minimal coupling of teleparallel gravity to a scalar field~\cite{TG-S}, and non-linear causality~\cite{OINC-IGO}. 

The conformal symmetry is very natural and fundamental symmetry of spacetime. 
It is a generalization of scale invariance where conformal factor depends on coordinates. 
It is known that conformal symmetry may be broken on quantum level. 
This leads to the famous conformal anomaly~\cite{Duff:1993wm}, 
which is known to be able to induce the inflationary cosmology (for a recent study, see~\cite{Hawking:2000bb} and also~\cite{Armillis:2013wya}). 
From another side, there exists well-known phenomena of asymptotic 
conformal invariance where at high energies the theory asymptotically 
approaches the conformal phase~\cite{Buchbinder:1992rb, Buchbinder:1985zx}. 
Furthermore, it is well-known that gravity theory may be conformally 
invariant (the Weyl tensor squared action) and pretend to be the candidate for 
a fundamental theory unifying all the four known fundamental interactions~\cite{Buchbinder:1992rb}.

Very recently, in Ref.~\cite{Bars:2013vba}, conformal symmetry has been 
studied in terms of the relation of cyclic cosmology and the metastability of 
the Higgs field. In addition, in Ref.~\cite{Kallosh:2013hoa}, a novel class of 
chaotic inflation models has been proposed, where conformal invariance is 
spontaneously broken. Therefore, currently, conformal symmetry of theories is a quite important issue in the literature. 
Furthermore, 
conformal transformation of $f(T)$ gravity~\cite{Yang:2010ji} 
and conformally-invariant teleparallel gravity~\cite{Maluf:2011kf} 
have also been investigated. 

In this paper, we concentrate on pure and extended teleparallel gravity and explore its conformal issues. We formulate conformal transformation 
in teleparallel gravity. In addition, we propose conformal scalar and gauge field theories and construct conformal torsion gravity. 
Moreover, we show that in extended teleparallel gravity with a conformal scalar field, a power-law or the de Sitter expansions of the universe can occur, 
and that in pure teleparallel gravity with a conformal scalar field, 
the $\Lambda$ cold dark matter ($\Lambda$CDM) model can be realized. 
Also, it is demonstrated that the de Sitter solution can exist in conformal torsion gravity,. 
We use units of $k_\mathrm{B} = c = \hbar = 1$ and denote the
gravitational constant $8 \pi G$ by
${\kappa}^2 \equiv 8\pi/{M_{\mathrm{Pl}}}^2$
with the Planck mass of $M_{\mathrm{Pl}} = G^{-1/2} = 1.2 \times 
10^{19}$\,\,GeV. 
Furthermore, the Heaviside-Lorentz units of electromagnetism is adopted. 

The paper is organized as follows. 
In Sec.\ II, we explain the fundamental formulations of teleparallel gravity. 
In Sec.\ III, we examine conformal transformation in teleparallel gravity. 
In addition, we propose conformal scalar and gauge field theories 
in Sec.\ IV and Sec.\ V, respectively. 
Moreover, we consider conformal torsion gravity in Sec.\ VI. 
Furthermore, in Sec.\ VII, we investigate cosmology in pure and extended teleparallel gravity with a conformal scalar field and in conformal torsion gravity. Finally, conclusions are presented in Sec.\ VIII. 

\section{Fundamental formulations of teleparallel gravity}\label{sec2}


Teleparallel gravity is described by using orthonormal tetrads components $e_A(x^{\mu})$, which are defined in the tangent space at each point $x^{\mu}$ of the manifold with the index $A$ running over $0, \dots, 3$. This allows us to rewrite the line element as follows 
\begin{eqnarray}
ds^{2} &=&g_{\mu\nu}dx^{\mu}dx^{\nu}=\eta_{ij}\theta^{i}\theta^{j}\label{1}\; ,\\
dx^{\mu}& =&e_{i}^{\;\;\mu}\theta^{i}\; , 
\quad 
\theta^{i}=e^{i}_{\;\;\mu}dx^{\mu}\label{2}\; ,
\label{1.1}
\end{eqnarray} 
where $\eta_{ij}=\text{diag}(-1,1,1,1)$ and $e_{i}^{\;\;\mu}e^{i}_{\;\;\nu}=\delta^{\mu}_{\nu}$ or $e_{i}^{\;\;\mu}e^{j}_{\;\;\mu}=\delta^{j}_{i}$. The square root of the metric determinant is given by $\sqrt{-g}=\det{\left[e^{i}_{\;\;\mu}\right]}=e$ and the tetrads $e^{a}_{\;\;\mu}$ represent the dynamic fields of the theory.
\par
In teleparallel gravity, the Weitzenb\"{o}ck connection is assumed instead of the Levi-Civita connection
\begin{eqnarray}
\tilde{\Gamma}^{\alpha}_{\mu\nu}=e_{i}^{\;\;\alpha}\partial_{\nu}e^{i}_{\;\;\mu}=-e^{i}_{\;\;\mu}\partial_{\nu}e_{i}^{\;\;\alpha}\label{co}\; .
\label{WC}
\end{eqnarray}
This leads to a null curvature but nonzero torsion. Hence, the main geometrical objects of the spacetime are constructed from this connection. The torsion tensor $T^{\alpha}_{\mu\nu}$ is defined as the antisymmetric part of the connection (\ref{WC})
\begin{eqnarray}
T^{\alpha}_{\;\;\mu\nu}&=&\tilde{\Gamma}^{\alpha}_{\nu\mu}-\tilde{\Gamma}^{\alpha}_{\mu\nu}=e_{i}^{\;\;\alpha}\left(\partial_{\mu} e^{i}_{\;\;\nu}-\partial_{\nu} e^{i}_{\;\;\mu}\right)\ , 
\label{tor}
\end{eqnarray}
while the components of the contorsion tensor $K^{\mu\nu}_{\;\;\;\;\alpha}$ are defined as
\begin{eqnarray}
K^{\mu\nu}_{\;\;\;\;\alpha}&=&-\frac{1}{2}\left(T^{\mu\nu}_{\;\;\;\;\alpha}-T^{\nu\mu}_{\;\;\;\;\alpha}-T_{\alpha}^{\;\;\mu\nu}\right)\ .
\label{contor}
\end{eqnarray}
The contorsion tensor basically refers to the difference between the Weitzenb\"{o}ck and Levi-Civita connections as is easily shown when an index is risen
\be
K^{\lambda}_{\;\; \mu\nu}= \tilde{\Gamma}^{\lambda}_{\mu\nu}-\Gamma^{\lambda}_{\mu\nu}=\frac{1}{2}\left(T_{\mu\;\;\ \nu}^{\;\; \lambda}+T_{\nu\;\;\ \mu}^{\;\; \lambda}-T_{\;\; \mu \nu}^{\lambda}\right)\; .
\label{contor2}
\ee
Thus, a new tensor $S_{\alpha}^{\;\;\mu\nu}$ can be defined by using the torsion and contorsion tensors
\begin{eqnarray}
S_{\alpha}^{\;\;\mu\nu}&=&\frac{1}{2}\left( K_{\;\;\;\;\alpha}^{\mu\nu}+\delta^{\mu}_{\alpha}T^{\beta\nu}_{\;\;\;\;\beta}-\delta^{\nu}_{\alpha}T^{\beta\mu}_{\;\;\;\;\beta}\right)\label{s}\;.
\label{ST}
\end{eqnarray}
The torsion scalar is constructed by contracting the indexes of the torsion scalar (\ref{tor}) with (\ref{s})
\begin{eqnarray}
T=T^{\alpha}_{\;\;\mu\nu}S^{\;\;\mu\nu}_{\alpha}\label{te}\; ,
\end{eqnarray}
which gives the well known action of teleparallel gravity~\cite{pereira1} 
\begin{eqnarray}\label{action}
S_G=\frac{1}{2\kappa^2}\int e\ T\ d^4x\; .
\label{TeleAction}
\end{eqnarray}
In addition, some generalizations of teleparallel gravity can be done, where the most simple leads to a generalization of the action (\ref{TeleAction}) to a more complex function of the torsion scalar (\ref{te}),  
\begin{eqnarray}\label{action}
S_G=\frac{1}{2\kappa^2}\int e\ f(T)d^4x\; .
\label{ftaction}
\end{eqnarray}
By supposing the existence of the matter action $S_\mathrm{m}=\int e \mathcal{L}_\mathrm{m}$ with 
$\mathcal{L}_\mathrm{m}$ matter Lagrangian and varying the action (\ref{ftaction}) 
with $S_\mathrm{m}=\int e \mathcal{L}_\mathrm{m}$ 
with respect to the tetrads, we get the following general equations of motion \cite{barrow,daouda1,daouda2}
\begin{eqnarray}
S^{\;\;\nu\rho}_{\mu}\partial_{\rho}Tf_{TT}+\left[e^{-1}e^{i}_{\mu}\partial_{\rho}\left(ee^{\;\;\alpha}_{i}S^{\;\;\nu\rho}_{\alpha}\right)+T^{\alpha}_{\;\;\lambda\mu}S^{\;\;\nu\lambda}_{\alpha}\right]f_{T}+\frac{1}{4}\delta^{\nu}_{\mu}f=\frac{\kappa^2}{2}\mathcal{T}^{\nu}_{\mu}\label{em}\; ,
\end{eqnarray}
where $\mathcal{T}^{\nu}_{\mu}$ is the energy momentum tensor of matter, $f_{T}=d f(T)/d T$ and $f_{TT}=d^{2} f(T)/dT^{2}$. Note that we have assumed $f(T)$ instead of $T$ in order to keep the most general action in terms of the torsion scalar $T$. These equations clearly depend on the choice of the set of tetrads, as shown in Ref.~\cite{Tamanini:2012hg}. In addition, the particular form of the action (\ref{ftaction}) can be reconstructed in order to reproduce the appropriate cosmological evolution (see Ref.~\cite{Rodrigues:2012qua}). 

Here, we are interested in studying the conformal transformation of this kind of theories and the conformal invariance of scalar fields and other matter 
fields. Accordingly, next sections are devoted to this analysis. Moreover, its comparison with curvature gravitational theories is a key role to understand this kind of torsion theories. In this sense, we can deduce the equivalence between GR and teleparallel gravity by writing the Ricci scalar in terms of the torsion tensor (see Ref.~\cite{pereira1}). Firstly, let us consider the Riemann tensor, which can be rewritten in terms of the contorsion tensor (\ref{contor2}) as follows
\bea
R^{\lambda}_{\;\;\mu\rho\nu}=\partial_{\rho}\Gamma^{\lambda}_{\mu\nu}-\partial_{\nu}\Gamma^{\lambda}_{\mu\rho}+\Gamma^{\lambda}_{\sigma\rho}\Gamma^{\sigma}_{\mu\nu}-\Gamma^{\lambda}_{\sigma\nu}\Gamma^{\sigma}_{\mu\rho}  \nn
=K^{\lambda}_{\;\;\mu\rho;\nu}-K^{\lambda}_{\;\;\mu\nu;\rho}+K^{\lambda}_{\;\;\sigma\nu}K^{\sigma}_{\;\;\mu\rho}-K^{\lambda}_{\;\;\sigma\rho}K^{\sigma}_{\;\;\mu\nu}\ .
\label{Riemann}
\eea
Here, the subscrip $_{;\mu}=D_{\mu}$ refers to the usual covariant derivative defined in terms of the Levi-Civita connection. Then, the Ricci tensor is easily obtained by contracting (\ref{Riemann})
\bea
R_{\mu\nu}=K^{\rho}_{\;\;\mu\rho;\nu}-K^{\rho}_{\;\;\mu\nu;\rho}+K^{\rho}_{\;\;\sigma\nu}K^{\sigma}_{\;\;\mu\rho}-K^{\rho}_{\;\;\sigma\rho}K^{\sigma}_{\;\;\mu\nu} \nn
=-S_{\nu\rho\mu}^{\;\;\;\;\;\; ;\rho}-g_{\mu\nu}T^{\sigma\;\;\; ;\rho}_{\;\;\rho\sigma} -S^{\rho\sigma}_{\;\;\;\mu}K_{\sigma\rho\nu}\ ,
\label{RT}
\eea
whereas the Ricci scalar is obtained by contracting the indexes in (\ref{RT}), which yields
\be
R=-T-2D^{\mu}T^{\nu}_{\;\;\mu\nu}\ .
\label{RS}
\ee
Hence, the Ricci scalar is equivalent to the torsion scalar plus a total derivative. By integration, the total derivative vanishes, so that the action of teleparallel gravity is completely equivalent to the Hilbert-Einstein action. 

\section{Conformal transformation in teleparallel gravity}

Conformal transformations are of quite interest in physics, since they can facilitate the analysis of some particular spacetimes, via the Penrose diagrams, or by connecting conformal frames where the treatment of the equations and degrees of freedom of a particular system turns out much easier. But it also may lead to an additional symmetry when conformal fields are involved, which may have interesting consequences both classical as quantum effects. In this section, we start by analyzing the conformal transformation of the geometrical objects defined in the previous section. Let us consider the conformal transformation
\begin{eqnarray}\label{ct}
\hat{g}_{\mu\nu}=\Omega^2(x)g_{\mu\nu},\ \ \hat{g}^{\mu\nu}=\Omega^{-2}(x)g^{\mu\nu}
\label{ConfT}
\end{eqnarray}
Then, the transformation of the tetrads fields is easily obtained by using (\ref{1.1}),
\begin{eqnarray}\label{ct1} 
\hat{e}^{a}_\mu=\Omega(x)e^{a}_\mu, \quad 
\hat{e}^{\mu}_a=\Omega^{-1}(x)e^{\mu}_a, \quad 
\hat{e}=\Omega^4e\ .
\end{eqnarray}
Under the conformal transformation (\ref{ct}), the torsion tensor is transformed as follows:
\begin{eqnarray}
\hat{T}^\rho_{\:\:\:\mu\nu}=T^\rho_{\:\:\:\mu\nu}+\Omega^{-1}[\delta^\rho_\nu \partial_\mu \Omega-\delta^\rho_\mu \partial_\nu \Omega]\ , 
\label{TC}
\end{eqnarray}
while the transformation of the contorsion tensor (\ref{contor}) yields
\be
\hat{K}^{\mu\nu}_{\;\;\;\rho}=\Omega^{-2}K^{\mu\nu}_{\;\;\;\rho}+\Omega^{-3}\left(\delta^{\nu}_{\rho}\partial^{\mu}\Omega-\delta^{\mu}_{\rho}\partial^{\nu}\Omega\ \right).
\label{KC}
\ee
Hence, by using the conformal transformations of the torsion tensor (\ref{TC}) and the contorsion tensor (\ref{KC}), the tensor (\ref{ST}) is reduced to 
\begin{equation}
\hat{S}_\rho^{\:\:\:\mu\nu}=\frac{1}{2}\Big(\hat{K}^{\mu\nu}_{\:\:\:\:\rho}+\hat{\delta}^\mu_
\rho \:\hat{T}^{\theta\nu}_{\:\:\:\:\theta}-\hat{\delta}^\nu_\rho\:
\hat{T}^{\theta\mu}_{\:\:\:\:\theta}\Big)=\Omega^{-2}S_\rho^{\:\:\:\mu\nu}+\Omega^{-3}(\delta^\mu_\rho \partial^\nu \Omega-\delta^\nu_\rho \partial^\mu \Omega)\ ,
\label{SC}
\end{equation}
and finally we obtain that the teleparallel Lagrangian is transformed into 
\begin{eqnarray}
\hat{T}=\Omega^{-2}T+4\Omega^{-3}g^{\mu\nu}\partial_\nu \Omega T^\rho_{\:\:\:\rho\mu}-6\Omega^{-4} g^{\mu\nu}\partial_\mu \Omega\partial_\nu \Omega\ ,
\end{eqnarray}
whereas the inverse transformation is given by
\begin{eqnarray}\label{T}
T \Eqn{=} \Omega^{2} \hat{T}-4\Omega^{-1}g^{\mu\nu}\partial_\nu \Omega T^\rho_{\:\:\:\rho\mu}+6\Omega^{-2} g^{\mu\nu}\partial_\mu \Omega\partial_\nu \Omega 
\nn 
\Eqn{=} \Omega^{2} \hat{T}-4\Omega \hat{g}^{\mu\nu}\partial_\nu \Omega \left(\hat{T}^\rho_{\:\:\:\rho\mu}+3\Omega^{-1}\partial_\mu \Omega\right)+6 \hat{g}^{\mu\nu}\partial_\mu \Omega\partial_\nu \Omega \nn 
\Eqn{=} \Omega^{2} \hat{T}-4\Omega \hat{g}^{\mu\nu}\partial_\nu \Omega \hat{T}^\rho_{\:\:\:\rho\mu}-6 \hat{g}^{\mu\nu}\partial_\mu \Omega\partial_\nu \Omega\ .
\label{TSC}
\end{eqnarray}
Note that here, we have used $ \hat{g}^{\mu\nu}=\Omega^{-2}(x)g^{\mu\nu}$, and the conformal transformation (\ref{TC}) for the term $T^{\rho}_{\;\;\;\rho\mu}$. Thus, with using these tools, we can investigate some conformal properties of extensions of teleparallel gravity. In the next section, we analyze some analogies with curvature gravity.

\section{Conformal scalar fields}

Firstly, let us briefly review conformal scalar fields in curvature gravity. We consider the action of the scalar field non-minimally coupling to gravity, 
given by
\be
S_{\phi}=\int d^4x \sqrt{-g}\left(\frac{1}{2}\nabla_{\mu}\phi\nabla^{\mu}\phi+\frac{A}{2} \phi^2 R -\frac{\phi^{n+1}}{n+1}\right)\ ,
\label{scr1}
\ee
where $R$ is the Ricci scalar, and $\{A,n\}$ are constants. By varying this action with respect to the scalar field $\phi$, the scalar field equation is obtained as
\be
\Box\phi-A\phi R+\phi^n=0\ .
\label{scr2}
\ee
We now define the conformal transformation of the scalar field as follows 
\be
\phi \equiv \e^{\beta\sigma}\hat{\phi}\ ,
\label{scr3}
\ee
where $\sigma=\sigma(\bf{x})$ and $\beta$ is a constant. 
It is known that 
an scalar field is called conformally invariant when the action (\ref{scr1}) and consequently the equation of motion (\ref{scr2}) are invariants under a conformal transformation of the metric (\ref{ConfT}) and the scalar field (\ref{scr3}). By redefining the conformal transformation (\ref{ConfT}) as $\Omega^2(x)=\e^{-\sigma(x)}$, the transformation of the metric tensor is given by $\hat{g}_{\mu\nu}=\e^{\sigma(x)}g_{\mu\nu}$, and by using the conformal transformation of the Ricci scalar (see for example Ref.~\cite{Faraoni:2006fx}), the scalar field equation (\ref{scr2}) is conformally invariant only if
\be
A=\frac{1}{6}\ , \quad \beta=-\frac{1}{2}\ , \quad n=3\ ,
\label{scr4}
\ee
which yields
\be
\Box\phi-A\phi R+\phi^n=
\left(\hat{\Box}\hat{\phi}-A\hat{\phi} \hat{R}+\hat{\phi}^n\right)\e^{(\beta-1)\sigma}\ .
\label{scr5}
\ee
This is a very well-known fact that has been widely studied in the literature (see Ref.~\cite{Buchbinder:1992rb} and references therein). Our proposal is to extend this analysis to teleparallel gravity by analogy with (\ref{scr2}), a scalar field non-minimally coupled to torsion can be considered, whose field equation, inspired by the curvature case (\ref{scr2}), could be written as follows
\be
\Box\phi-B\phi T+\phi^m=0\ ,
\label{scalareq}
\ee
where $\{B,m\}$ are constants. By assuming the conformal transformation of the scalar field (\ref{scr3}) again, where $\beta$ must be determined by imposing conformal invariance, and the transformation of the torsion scalar (\ref{TSC}), where $\Omega^2(x)=\e^{-\sigma(x)}$, the scalar field equation (\ref{scalareq}) is conformally transformed into 
\bea
&&
\left\{\hat{\Box}\phi-B\hat{\phi} \hat{T}+\e^{\left[\beta(m-1)+1\right]\sigma}\hat{\phi}^m+\beta\hat{\phi}\hat{\Box}\sigma+\hat{\phi}\left[\beta(\beta+1)+Ê\frac{3}{2}B\right]\partial_{\mu}\sigma\partial^{\mu}\sigma+(2\beta+1)\partial_{\mu}\sigma\partial^{\mu}\hat{\phi}-2B\hat{\phi}\partial^{\mu}\sigma\hat{T}^{\rho}_{\:\:\:\:\rho\mu}\right\}
\nonumber \\
&&
{}\times \e^{(\beta-1)\sigma}=0\ .
\label{sc2}
\eea
In analogy with the curvature case, we may assume the values (\ref{scr4}), where now turns out $B=\frac{1}{6}$, $\beta=-\frac{1}{2}$ and $m=3$, the scalar field equation (\ref{sc2}) reads
\be
\hat{\Box}\phi-\frac{1}{6}\hat{\phi} \hat{T}+\hat{\phi}^m+\frac{1}{2}\hat{\phi}\hat{\Box}\sigma-\frac{1}{3}\hat{\phi}\partial^{\mu}\sigma\hat{T}^{\rho}_{\:\:\:\:\rho\mu}=0\ .
\label{sc3}
\ee 
As shown, the scalar field equation is not conformally invariant, but two additional terms that are not present in the case of curvature gravity, remain in the equation. 
In fact, there is no combination of values for the parameters $\{\beta,B,m\}$ that leads to an invariant  scalar field equation (\ref{scalareq}). It is straightforward to show that the reason lies in the fact that teleparallel gravity, described by the action (\ref{TeleAction}) is completely equivalent to GR up to a total derivative, as shown previously, such that no extension including non-minimally couplings or general functions $f(T)$ are equivalent to the analog theory in curvature gravity due to the presence of the total derivative in (\ref{RS}). Nevertheless, following the relation (\ref{RS}), one may consider a non-minimally coupled scalar field described by the action~\cite{Maluf:2011kf}
\be
S_{\phi}=\int d^4x\ e\left[\frac{1}{2}\nabla_{\mu}\phi\nabla^{\mu}\phi-\frac{1}{2}\phi^2\left(C T+D\nabla^{\mu}T^{\rho}_{\;\;\;\mu\rho}\right)-\frac{\phi^{m+1}}{m+1}\right]\ .
\label{sca}\ee
Here, recall that $e=\sqrt{-g}$, and $\{C,D,\beta,m\}$ are free parameters. Then, by varying the action with respect to the scalar field $\phi$, the field equation yields
\be
\Box\phi+C\phi T+D\phi\nabla^{\mu}T^{\rho}_{\;\;\;\mu\rho}+\phi^m=0\ .
\label{scb} 
\ee
Alternatively, one might integrate the third term in (\ref{sca}) by parts, so the action (\ref{sca}) yields
\be
S_{\phi}=\int d^4x\ e\left[\frac{1}{2}\nabla_{\mu}\phi\nabla^{\mu}\phi-\frac{C}{2}\phi^2T+D\phi T^{\rho}_{\;\;\;\mu\rho}\nabla^{\mu}\phi-\frac{\phi^{m+1}}{m+1}\right]\ ,
\label{sca2}\ee
which also leads to the field equation for the scalar field (\ref{scb}) while varying the action (\ref{sca2}) with respect to $\phi$. Then, by applying the conformal transformations obtained in the previous section and the transformation for the scalar field (\ref{scr3}), the equation (\ref{scb}) is transformed into 
\[
\left\{\hat{\Box}\phi+C\hat{\phi} \hat{T}+D\hat{\phi}\hat{\nabla}^{\mu}\hat{T}^{\rho}_{\;\;\;\mu\rho}+\e^{\left[\beta(m-1)+1\right]\sigma}\hat{\phi}^m+\left(\beta+\frac{3}{2}D\right)\hat{\phi}\hat{\Box}\sigma+\hat{\phi}\left[\beta(\beta+1)-Ê\frac{3}{2}(C-D)\right]\partial_{\mu}\sigma\partial^{\mu}\sigma \right. 
\]
\be
\left. +(2\beta+1)\partial_{\mu}\sigma\partial^{\mu}\hat{\phi}+(2C-D)\hat{\phi}\partial^{\mu}\sigma\hat{T}^{\rho}_{\:\:\:\:\rho\mu}\right\}\e^{(\beta-1)\sigma}=0\, .
\label{sc2b}
\ee
It is straightforward to show that the extra terms appearing in (\ref{sc2b}) after the conformal transformation can be removed and the scalar field equation (\ref{scb}) turns out conformally invariant, when the set of free parameters $\{C,D,\beta,m\}$ are taken as 
\be
C=\frac{1}{6}\ , \quad D=\frac{1}{3}\ , \quad \beta=-\frac{1}{2}\ , \quad m=3\ .
\label{scc}
\ee
Then, the original equation (\ref{scb}) is recovered, and the conformal transformation defined above becomes a symmetry of the model 
\be
\left(\hat{\Box}\hat{\phi}+\frac{1}{6}\hat{\phi} \hat{T}+\frac{1}{3}\hat{\phi}\hat{\nabla}^{\mu}\hat{T}^{\rho}_{\;\;\;\mu\rho}+\hat{\phi}^3\right)\e^{-3\sigma/2}=0\ .
\label{scd} 
\ee
Hence, the scalar field equation remains invariant. This emphasizes the relation between curvature and torsion (\ref{RS}), which is the total derivative and a key ingredient for reconstructing the conformal invariant scalar field model (\ref{scb}). 

Here, it would again be significant to stress the novel ingredient obtained in this section. 
In the framework of teleparallelism constructed with the Weitzenb\"{o}ck connection, the crucial condition for the torsion scalar $T$ to have 
its non-minimal coupling to a scalar field in a conformal manner 
is that the Ricci scalar consisting of the torsion scalar and 
the total derivative of the torsion tensor $T^{\alpha}_{\mu\nu}$ 
as shown in Eq.~(\ref{RS}) 
has to be involved in the action of the scalar field in Eq.~(\ref{sca}). 
It is considered that this consequence would be reasonable 
because not only the torsion scalar $T$ but also the Ricci scalar $R$ 
are taken into consideration. 
This is our new result, which has first been noticed in the present work and 
not been studied yet in the past works.

Furthermore, it should be important to explore whether a pure $f(T)$ gravity without any scalar field might be  conformally invariant  
or not. Provided that even in teleparallelism, the conformal transformation 
in the ordinary curvature gravity can be considered. 
When we regard the action in Eq.~(\ref{sca}) as a kind of  action written
in the so-called Jordan frame, by making the conformal transformation from 
the Jordan frame to the Einstein frame,  the transformed action is obtained in the 
Einstein frame. However, from the form of the action 
in Eq.~(\ref{sca}), it can presumably be considered that there exist 
the term of the pure torsion scalar term and some potential term consisting of 
the conformal scalar field in the resultant action in the Einstein frame, similarly to that in ordinary curvature gravity. Thus, it follows from these investigations that a pure $f(T)$ gravity without any scalar fields might not be conformally invariant. 

\section{Gauge fields}

Let us consider now the case of a gauge field described by the action
\be
S_{F}=\int d^4x\ e \left(-\frac{1}{4}F_{\mu\nu}F^{\mu\nu}\right)\ ,
\label{gf1}
\ee
where, as usual, 
\be
F_{\mu\nu}=D_{\mu}A_{\nu}-D_{\nu}A_{\mu}\ . 
\label{gf2}
\ee
Note that the model minimally coupling to gravity, which is described by the action (\ref{gf1}), can trivially be extended to teleparallel gravity, since in this case there is no coupling between the gauge field and the torsion scalar, and hence that as in curvature gravity, the Lagrangian is conformally invariant because the energy-momentum tensor is traceless. Consequently, by assuming that $\hat{A}_{\mu}=A_{\mu}$, the transformation of the action (\ref{gf1}) yields
\be
\hat{F}_{\mu\nu}=\hat{D}_{\mu}A_{\nu}-\hat{D}_{\nu}A_{\mu}=F_{\mu\nu}\ .
\label{gf3}
\ee
Furthermore, the action (\ref{gf1}) is transformed into 
\be
\hat{S}_{F}=\int d^4x\ \hat{e}\ \Omega^{-4} \left[-\frac{1}{4}\left(\Omega^2\hat{g}^{\mu\lambda}\right)\left(\Omega^2\hat{g}^{\nu\rho}\right)F_{\mu\nu}F_{\lambda\rho}\right]=\int d^4x\ \hat{e} \left(-\frac{1}{4}\hat{F}_{\mu\nu}\hat{F}^{\mu\nu}\right)=S_F\ .
\label{gf4}
\ee
However, note that this is the simplest case and a trivial extension from curvature gravity, but when non-minimal couplings to gravity are taken into account, the action will not be conformally invariant in general. 

\section{Conformal torsion gravity}
Following the previous sections, this section is devoted to the reconstruction of an analogy to the Weyl theory in curvature gravity. In this sense, the following tensor is defined as 
\be
C^{\rho}_{\;\;\mu\nu}=T^{\rho}_{\;\;\mu\nu}+S^{\rho}_{\;\;\mu\nu}\ , 
\label{CT1}
\ee
where $T^{\rho}_{\mu\nu}$ is the torsion tensor (\ref{tor}) and $S^{\rho}_{\;\;\mu\nu}$ is the tensor defined in (\ref{ST}). It is straightforward to show that the tensor (\ref{CT1}) is conformally invariant under the transformations (\ref{TC}) and (\ref{SC})
\be
\hat{C}^{\rho}_{\mu\nu}=\hat{T}^{\rho}_{\mu\nu}+\hat{S}^{\rho}_{\mu\nu}=T^{\rho}_{\mu\nu}+\Omega^{-1}\left[\delta^{\rho}_{\nu}\partial_\mu\Omega-\delta^{\rho}_{\mu}\partial_\nu\Omega\right]+S^{\rho}_{\mu\nu}+\Omega^{-1}\left[\delta^{\rho}_{\mu}\partial_\nu\Omega-\delta^{\rho}_{\nu}\partial_\mu\Omega\right]=T^{\rho}_{\mu\nu}+S^{\rho}_{\mu\nu}=C^{\rho}_{\mu\nu}\ .
\label{CT2}
\ee
Therefore, the following action may be considered
\be
S=\frac{1}{2\kappa^2}\int\ d^4x\ e\ C^{\rho}_{\;\;\mu\nu} C_{\rho}^{\; \; \mu\nu}\ .
\label{CT3}
\ee
Nevertheless, note that this action is not conformally invariant, because $\hat{C}^{\rho}_{\;\;\mu\nu} \hat{C}_{\rho}^{\; \; \mu\nu}=\Omega^{-2} C^{\rho}_{\;\;\mu\nu} C_{\rho}^{\; \; \mu\nu}$, but $\hat{e}=\Omega^4\ e$, which yields
\be
\int\ d^4x\ \hat{e}\ \hat{C}^{\rho}_{\;\;\mu\nu} \hat{C}_{\rho}^{\; \; \mu\nu}=\int\ d^4x\ \Omega^2\ e\ C^{\rho}_{\;\;\mu\nu} C_{\rho}^{\; \; \mu\nu}\ .
\label{CT4}
\ee
However, a second power of the action (\ref{CT4}) becomes conformally invariant
\be
\int\ d^4x\ \hat{e}\ \left[\hat{C}^{\rho}_{\;\;\mu\nu} \hat{C}_{\rho}^{\; \; \mu\nu}\right]^2=\int\ d^4x\  e\left[\ C^{\rho}_{\;\;\mu\nu} C_{\rho}^{\; \; \mu\nu}\right]^2\ .
\label{CT5}
\ee 
Moreover, the scalar $\mathcal{C} \equiv C^{\rho}_{\;\;\mu\nu} C_{\rho}^{\; \; \mu\nu}$ can be written in terms of the torsion tensor and torsion scalar as follows
\be
\mathcal{C}=C^{\rho}_{\;\;\mu\nu} C_{\rho}^{\; \; \mu\nu}=\frac{1}{8}T_{\mu\nu}^{\;\;\;\;\rho}T^{\;\;\nu\mu}_{\rho}+\frac{19}{16}T_{\mu\nu}^{\;\;\;\;\rho}T^{\mu\nu}_{\;\;\;\;\rho}+\frac{1}{2}T_{\;\;\nu\alpha}^{\alpha}T^{\beta\nu}_{\;\;\;\;\beta}+2T\ .
\label{CT6}
\ee
Thus, in comparison with the Weyl gravity, a conformally-invariant action in the framework of torsion gravity can be constructed by starting from the tensor (\ref{CT1}).

Finally, 
we mention the following two points regarding the conformal torsion gravity 
built with the tensor defined in Eq.~(\ref{CT1}). 
First, note that the Minkowski spacetime is a vacuum solution of this theory, since all the terms in (\ref{CT6}) becomes null for a flat spacetime. Second, in analogy to the Weyl gravity, the scalar (\ref{CT6}) is null for a FLRW metric, so that there is no homogeneous and isotropic cosmological solutions in the framework of a theory described by the action (\ref{CT5}), as occurs in the well known case of the Weyl gravity. Hence, this theory is not equivalent to the Einstein gravity, as can be also shown by substituting  the relations between the torsion magnitudes and the curvature terms given by the expressions (\ref{Riemann}-\ref{RS}), into the action (\ref{CT5}). By doing so, some extra terms, which are suppressed in the usual teleparallel gravity, remain in the action (\ref{CT5}), leading to a different action than the Hilbert-Einstein action.

\section{Field equations and cosmological solutions}

In this section, the above reconstructed models, which present a conformal symmetry, are considered and the corresponding field equations are obtained. Here, we mainly focus on the flat Friedmann-Lema\^{i}tre-Robertson-Walker 
(FLRW) metric
\be
ds^2=dt^2-a^2(t)\sum_{i=1}^{3}(dx^i)^2\ .
\label{FLRW}
\ee
A diagonal set of tetrads is assumed, which for the above metric can be expressed as follows
\be
e^{i}_{\;\;\mu}=\text{diag}\left(1,a,a,a\right)\ , \quad e_{i}^{\;\;\mu}=\text{diag}\left(1,\frac{1}{a},\frac{1}{a},\frac{1}{a}\right)\ .
\label{FLRWtetrad}
\ee
It is straightforward to check that (\ref{FLRWtetrad}) corresponds to the metric (\ref{FLRW}) trough the relation (\ref{1.1}). The determinant of the matrix is $e=a^3$, whereas the components of the torsion tensor (\ref{tor}) and contorsion tensor (\ref{contor2}) for the tetrads (\ref{FLRWtetrad}) are given by
\begin{eqnarray}
T^{1}_{\;\;01}=T^{2}_{\;\;02}=\,T^{3}_{\;\;03}=K^{01}_{\;\;\;\;1}=K^{02}_{\;\;\;\;2}=K^{03}_{\;\;\;\;3}=H(t)\; ,\label{torsiontype3}
\end{eqnarray}
where the Hubble parameter is defined by $H\equiv\dot{a}/a$ as usual. The components of the tensor $S_{\alpha}^{\;\;\mu\nu}$ in (\ref{s}) yield
\begin{eqnarray}
S_{1}^{\;\;10}=S_{2}^{\;\;20}=S_{3}^{\;\;30}=H(t)\,\;.\label{tensortype3}
\end{eqnarray}
Hence, by using (\ref{torsiontype3}) and (\ref{tensortype3}), the torsion scalar (\ref{te}) is given by
\begin{eqnarray}
T=-6H^2(t)\; \label{torsionScalar1}.
\end{eqnarray}
 
For this class of spacetimes, and by the choice of the tetrads (\ref{FLRWtetrad}), some cosmological solutions are reconstructed, where the occurrence of a late-time accelerating phase is obtained. First, the scalar field described by the action (\ref{sca2}) is considered in the framework of $f(T)$ gravities, and the FLRW equations are deduced and the Lagrangian $f(T)$ is reconstructed for some kind of cosmological solutions. Next, the field equations of conformal torsion gravity (\ref{CT5}) are obtained, and the FLRW spacetimes are analyzed.

\subsection{Conformal scalar field theory}

Let us start considering the $f(T)$ action (\ref{ftaction}) and the conformal invariant scalar field (\ref{sca2})
\be
S=\int d^4x e\left[f(T)+\frac{1}{2}\nabla_{\mu}\phi\nabla^{\mu}\phi-\frac{C}{2}\phi^2T+D\phi T^{\rho}_{\;\;\mu\rho}\nabla^{\mu}\phi-\frac{\phi^{m+1}}{m+1}+2\kappa^2\mathcal{L}_\mathrm{m}\right]\ .
\label{FEQ1}
\ee 
By varying the action (\ref{FEQ1}) with respect to the tetrad, the field equations yield
\bea
&&
S_{\mu}^{\;\;\nu\rho}f_{TT}\partial_{\rho}T+\left[e^{-1}e^{i}_{\;\;\mu}\partial_{\rho}\left(ee^{\;\;\alpha}_{i}S^{\;\;\nu\rho}_{\alpha}\right)+T^{\alpha}_{\;\;\lambda\mu}S^{\;\;\nu\lambda}_{\alpha}\right]f_{T}+\frac{1}{4}\delta^{\nu}_{\mu}f\nn 
&&
{}= \frac{\kappa^2}{2}\mathcal{T}^{\nu}_{\mu}+C\left[\left(\frac{1}{2}T^{\sigma}_{\;\;\rho\mu}S_{\sigma}^{\;\;\nu\rho}+\frac{1}{8}\delta^{\nu}_{\mu}T\right)\phi^2+\frac{1}{2}e^{-1}e_{\;\;\mu}^{i}\partial_{\sigma}\left(\phi^2e e^{\;\;\rho}_{i}S_{\rho}^{\;\;\nu\sigma}\right)\right] \nn 
&&
{}+\frac{1}{4}\partial_{\mu}\phi\partial^{\nu}\phi-\frac{1}{8}\delta_{\mu}^{\nu}\partial_{\rho}\phi\partial^{\rho}\phi-\frac{D}{4}\left[\phi\partial^{\rho}\phi\left(\delta^{\nu}_{\mu}T^{\alpha}_{\;\;\rho\alpha}-T^{\nu}_{\;\;\sigma\mu}\right)-\phi T^{\alpha}_{\;\;\rho\alpha}\left(\delta^{\rho}_{\mu}\partial^{\nu}\phi+g^{\rho\nu}\partial_{\mu}\phi\right)\right. \nn 
&& 
\left. 
{}-e^{-1}e_{\;\;\mu}^{i}\partial_{\sigma}\left(\phi e e^{\;\;\nu}_{i}\partial^{\sigma}\phi-\phi e e^{\;\;\sigma}_{i}\partial^{\nu}\phi\right)\right]+\delta^{\nu}_{\mu}\frac{\phi^{m+1}}{4(m+1)}\ .
\label{FEQ2}
\eea 
For the flat FLRW metric (\ref{FLRW}), the Friedmann equation reads 
\be
3H^2f_T+\frac{1}{4}f-\frac{3}{4}H\phi\left(CH\phi+D\dot{\phi}\right)-\frac{1}{8}\dot{\phi}^2-\frac{\phi^{m+1}}{4(m+1)}-\frac{1}{2}\kappa^2\rho_\mathrm{m}=0\ ,
\label{FQE3}
\ee
whereas the scalar field equation (\ref{scb}) is given by
\be
\ddot{\phi}+3H\dot{\phi}-6CH^2\phi+3D\left(\dot{H}+3H^2\right)\phi+\phi^m=0\ .
\label{FQE4}
\ee
Firstly, we might consider cosmological solutions of the type
\be
H=\frac{\alpha}{t}\ ,
\label{pwl1}
\ee
that would be referred to as power-law behavior. 
Here, $\alpha (>0)$ is a positive constant for $t>0$. 
Within GR, this type of solutions accomplishes the scale factor evolution for perfect fluids with a constant equation of state (EoS), such as the dust ($\alpha = 2/3$) or radiation ($\alpha = 1/2$) dominated stage of the universe. Moreover, $\alpha>1$ gives rise to an accelerating expansion, which could be related to the inflationary phase as well as the dark energy epoch. Note that in general the set of equations (\ref{FQE3})-(\ref{FQE4}) can not be solved exactly, and even for simple solutions as (\ref{pwl1}). Nevertheless, in absence of the scalar potential, the scalar field equations remains conformal invariant, and both equations turn out much simpler. Therefore, for the class of solutions (\ref{pwl1}), the scalar field equation (\ref{FQE4}) in absence of potential reduces to
\be
t^2 \ddot{\phi}+3\alpha t\dot{\phi}+\alpha\left(2\alpha-1\right)\phi=0\ ,
\label{pwl2}
\ee
where we have assumed the values of the free parameters (\ref{scc}) that lead to a conformally-invariant scalar field equation, (\ref{FQE4}). Hence, the equation (\ref{pwl2}) can be easily solved, leading to
\be
\phi(t)=k_1t^{-\alpha}+k_2t^{1-2\alpha}\ .
\label{pwl3}
\ee
Here, $k_i$ are integration constants. The corresponding $f(T)$ is reconstructed so that the Friedmann equation (\ref{FQE3}) can be satisfied. In the absence of any kind of matter, the action $f(T)$ yields
\be
f(T)=\tilde{k}_1\left(-\frac{T}{\alpha^2}\right)^{2\alpha}+k_3\sqrt{-T}\ , \quad \text{where} \quad \tilde{k}_1=-\frac{2^{-2(1+\alpha)}9^{-\alpha}(\alpha-1)^2}{2\alpha-1/2}k_1\ ,
\label{pwl4}
\ee
and $k_3$ is an integration constant. In case that a particular perfect fluid is assumed, whose equation of state is described by $p_\mathrm{m}=w_\mathrm{m}\rho_\mathrm{m}$, and satisfies the continuity equation
\be
\dot{\rho}_\mathrm{m}+3H(1+w_\mathrm{m})\rho_\mathrm{m}=0\ .
\label{pwl5}
\ee
The solution of (\ref{pwl5}) for the class of Hubble parameters (\ref{pwl1}) yields
\be
\rho_\mathrm{m}=\rho_0t^{-3\alpha(1+w_\mathrm{m})}=\rho_0\left(-\frac{T}{6\alpha^2}\right)^{\frac{3\alpha(1+w_\mathrm{m})}{2}}\ .
\label{pwl6}
\ee
Accordingly, by inserting this result into (\ref{FQE3}), the equation turns out a differential equation on $T$, which leads to
\be
f(T)=\tilde{k}_1\left(-\frac{T}{\alpha^2}\right)^{2\alpha}+k_3\sqrt{-T}+\rho_0\frac{4}{1-3\alpha(1+w_\mathrm{m})}\left(-\frac{T}{6\alpha^2}\right)^{\frac{3\alpha(1+w_\mathrm{m})}{2}}\ .
\label{pwl7}
\ee
Hence, the class of solutions (\ref{pwl1}) can be reproduced in the framework of $f(T)$ gravities with the presence of a conformal scalar field. Note that the case of radiation-like expansion ($\alpha=1/2$), gives rise to a different solution than (\ref{pwl4}) and (\ref{pwl7}), but also in terms of powers of the torsion scalar $T$. 

Let us consider now another natural and important solution in cosmology, the de Sitter evolution, which depicts an exponential growth of the expansion, and is believed to be approximately the behavior of dark energy today as well as 
that of inflation in the early universe
\be
H=H_0\ .
\label{dS}
\ee
In such a case, the scalar field equation (\ref{FQE4}) becomes
\be
\ddot{\phi}+3H_0\dot{\phi}+2H_0^2\phi+V_0\phi^3=0\ ,
\label{dS2}
\ee
where the scalar potential has been rewritten as $V(\phi)=V_0\phi$ for convenience. Note that for a constant Hubble parameter (\ref{dS}), the term $3H_0^2f_T+\frac{1}{4}f$ in (\ref{FQE3}) is time independent, so the rest of the equation should also be time independent, which leads to the constraint
\be
\frac{1}{4}H_0\phi\left(\frac{1}{2}H_0\phi+\dot{\phi}\right)+\frac{1}{8}\dot{\phi}^2+\frac{V_0}{16}\phi^{4}=K\ ,
\label{dS3}
\ee
where $K$ is a constant, and the Friedmann equation (\ref{FQE3}) is reduced to an algebraic equation
\be
3H_0^2f_T(T_0)+\frac{1}{4}f(T_0)+K=0\ .
\label{dS4}
\ee
Recalling that $T_0=-6H_0^2$, the solutions of the algebraic equation (\ref{dS4}) represent different de Sitter points, which may be identified with the accelerating epochs of the expansion history. The most simple solution is given by a constant scalar field $\phi(t)=\phi_0$, which under the equation (\ref{dS2}) yields
\be
\phi_0=\pm4H_0\sqrt{\frac{2}{|V_0|}}\ ,
\label{dS5}
\ee
where $V_0<0$ is assumed in order to avoid a complex scalar field. Therefore, $K=\frac{60H_0^4}{|V_0|}$ in (\ref{dS3}), and for a particular $f(T)$  if the equation (\ref{dS4}) contains real and positive roots, a de Sitter expansion exists as a particular solution of the Friedmann equations. Furthermore, the equation (\ref{dS3}) contains also time-dependent solutions for the scalar field, which are compatible with the scalar field equation (\ref{dS2}). In this sense, for a null potential $V_0=0$ and $K=0$, the solution of the constraint equation (\ref{dS3}) is given by
\be
\phi(t)=k_1\e^{-H_0t}\ ,
\label{dS6}
\ee
which also satisfies the scalar field equation (\ref{dS2}). Thus, the algebraic equation (\ref{dS4}) reduces to $3H_0^2f_T(T_0)+\frac{1}{4}f(T_0)=0$, whose positive and real roots $H_0$ represent de Sitter solutions for this model. Hence, by a particular Hubble parameter, the corresponding action $f(T)$ can be reconstructed. Moreover, the $\Lambda$CDM model is also studied with the presence of the conformal scalar field below.

\subsection{Conformal torsion gravity}

Let us recall $\mathcal{C}=C^{\rho}_{\;\;\mu\nu} C_{\rho}^{\; \; \mu\nu}$. The field equations for the action (\ref{CT5}) are obtained by varying the Lagrangian with respect to the tetrads
\bea
\mathcal{C}\left[\frac{1}{2}\left(T^{\rho\sigma\nu}T_{\sigma\rho\mu}-T^{\nu\rho\sigma}T_{\rho\sigma\mu}\right)-\frac{19}{4}T^{\rho\sigma\nu}T_{\rho\sigma\mu}-T^{\rho}_{\;\;\mu\rho}T^{\sigma\nu}_{\;\;\;\;\sigma}-T^{\rho\alpha}_{\;\;\;\;\rho}T^{\nu}_{\;\;\alpha\mu}+8T^{\rho}_{\;\;\alpha\mu}S_{\rho}^{\;\;\nu\alpha}\right] \nn
-e^{-1}e^{i}_{\;\;\mu}\partial_{\rho}\left\{e\mathcal{C}\left[e_{i}^{\;\;\sigma}\left(\frac{1}{4}T^{\nu\rho}_{\;\;\;\;\sigma}-\frac{1}{4}T^{\rho\nu}_{\;\;\;\;\sigma}+\frac{19}{4}T^{\;\;\;\;\rho\nu}_{\sigma}-8S_{\sigma}^{\;\;\nu\rho}\right)+e_{i}^{\;\;\nu}T^{\beta\rho}_{\;\;\;\;\beta}-e_{i}^{\;\;\rho}T^{\beta\nu}_{\;\;\;\;\beta}\right]\right\}+\frac{1}{2}\delta^{\nu}_{\mu}\mathcal{C}^2=\kappa^2\mathcal{T}^{\nu}_{\mu}\ .
\label{FQE5}
\eea 
Note that $\mathcal{C}=0$ for an isotropic and homogeneous spacetime as the FLRW metric (\ref{FLRW}) and the set of tetrads (\ref{FLRWtetrad}), in analogy to the Weyl tensor in curvature gravity, which is also null within this class of spacetimes. Hence, the left-hand side (l.h.s.) of the field equations (\ref{FQE5}) becomes null, and there is no FLRW solution for the conformal action (\ref{CT5}). Furthermore, any action proportional to $\mathcal{C}^n$ $(n>1)$, is not compatible with FLRW spacetimes, unless other geometrical terms are included, as some function  $f(T)$, which would reduce the FLRW equations to the usual $f(T)$ equations. 

Finally, we remark that $\mathcal{C}$ gives no contribution to the expansion of the universe, and therefore that $\mathcal{C}$, which is a conformal invariant 
scalar, plus the conformal invariant scalar field can lead to some solution describing the FLRW universe, namely, scalar part makes contribution to cosmology. 
In addition, by making analogy with the Weyl-squared gravity $\mathcal{C}^2$~\cite{Mannheim:2005bfa}, it is also found that $\mathcal{C}^2$ does not give any contribution to the FLRW cosmology but black hole solutions may change.

\subsection{$\Lambda$CDM model}

Let us now explore the $\Lambda$CDM model in the framework of the theories studied above. The Hubble parameter of the $\Lambda$CDM model can be written as
\be
H=\frac{\kappa}{3}\rho_{\mathrm{m} 0}
a^{-3}+\frac{\Lambda}{3}=\frac{\kappa}{3}\rho_{\mathrm{m} 0}
(1+z)^3+\frac{\Lambda}{3}\ ,
\label{LCDM1a}
\ee
where $\rho_{\mathrm{m} 0}$ is the present matter density and 
$\Lambda$ is a constant. 
Furthermore, in deriving the second equality in (\ref{LCDM1a}) we have used 
the relation $a_0/a = 1+z$ with $a_0 = 1$, where $a_0$ is the current value of the scale factor and $z$ is the redshift. 
This solution has widely been explored in the literature in the framework of modified gravities, where it was found that might be reproduced in the absence of a cosmological constant (see~\cite{delaCruzDombriz:2006fj}). By focusing on the conformal scalar field considered above, the corresponding $f(T)$ action might be reconstructed from the Friedmann equation (\ref{FQE3}) and the scalar field equation (\ref{FQE4}). Nevertheless, in general it is not possible to get an exact solution for the action $f(T)$ because the equations are highly non linear and numerical resources are required. However, by using some approximations and a qualitative analysis, it can be shown that the $\Lambda$CDM model (\ref{LCDM1a}) can be reproduced in the context of teleparallel gravity with the presence of a conformally-invariant scalar field and with no need of a cosmological constant. Therefore, note that for a general Hubble parameter $H(t)$, and that by assuming a null scalar potential, $V_0=0$, a general solution of the scalar field equation (\ref{FQE4}) yields
\be
\phi=\exp\left[-\int^t_{t_0} H(t') dt'\right]\ ,
\label{LCDM1}
\ee
which also satisfies (\ref{dS3}) when $H=H(t)$ and $K=0$. Hence, the Friedmann equation (\ref{FQE3}) reduces to the one in $f(T)$ gravity, and by setting $f(T)=T$, to the Friedmann equation of teleparallel gravity
\be
H^2=\frac{\kappa^2}{3}\rho_\mathrm{m} \ .
\label{LCDM2}
\ee
Thus, the matter-dominated epoch is realized by setting $V(\phi)\sim0$ along that phase. Later on, when the Hubble parameter $H(t)\sim H_0$, the solution (\ref{dS5}) may be recovered, where now $V(\phi)<0$, and the scalar field becomes constant. As a result, the late-time acceleration epoch is reproduced, leading to an effective cosmological constant $\Lambda=\frac{120 H_0^4}{|V_0|}$ as found above, which provides a well description of the $\Lambda$CDM model in the absence of any cosmological constant but with the presence of a conformally-invariant scalar field. 

In addition, a particular non null case from the conformal torsion gravity investigated previously is the one given by a linear action on $\mathcal{C}$ (\ref{CT3}), whose Friedmann equation becomes 
\be
H^2=\frac{8\kappa^2}{9}\rho_\mathrm{m}\ .
\label{FQE6}
\ee
Hence, by the appropriate definition of the coupling constant $\kappa^2$, the Friedmann equation (\ref{FQE6}) reduces to the case of teleparallel gravity, or equivalently to GR. By including a cosmological constant and the right expression for the coupling constant $\kappa^2$, the Friedmann equation reads
\be  
H^2=\frac{8\pi G}{3}\rho_{\mathrm{m}}+\frac{\Lambda}{3}\ .
\label{FQE7}
\ee
Consequently, the linear action in $\mathcal{C}$ turns out to be the usual Friedmann equation, where the $\Lambda$CDM model can be reproduced. Moreover, the cosmological constant can be substituted by the conformal scalar field, which reduces to the case explored above, where it was shown qualitatively that the $\Lambda$CDM model can be reproduced. Thus, an action given by $\mathcal{C}$ with the presence of a conformal scalar field is capable of reproducing the $\Lambda$CDM model.

\section{Conclusions}

In the present paper, we have examined some conformal issues of pure and extended teleparallel gravity. We have formulated the conformal transformation of teleparallel gravity. Also, we have proposed conformal scalar and gauge field theories and constructed conformal torsion gravity. 
It has explicitly been shown that in extended teleparallel gravity with a conformal scalar field, a power-law or the de Sitter expansions of the universe can 
be realized, and that pure teleparallel gravity with a conformal scalar field 
can give rise to the $\Lambda$CDM model. 
Furthermore, it has been demonstrated that in conformal torsion gravity, the de Sitter solution can be realized. 

By imposing conformal invariance, a particular scalar field model that exhibits this class of symmetry, has been reconstructed. Note that in comparison with curvature gravity some additional terms are included in the action. In particular, a derivative of the torsion tensor has to be assumed in the action in order to satisfy conformal invariance, due to the presence of a total derivative in the relation among the curvature scalar and the torsion scalar (\ref{RS}). The case of gauge fields is much more straightforward since the energy-momentum tensor is traceless, which automatically ensures the conformal invariance of the field equation for the gauge field, as in curvature gravity. Moreover, a conformal tensor based on combinations of the torsion tensor and scalar has been found, which leads to an invariant conformal torsion gravity. As in the case of the Weyl gravity, the scalar $\mathcal{C}$ becomes null in homogeneous and isotropic spacetimes (FLRW), and then the Friedmann equation turns out null. Nevertheless, we have shown that for a linear action, the first Friedmann equation reduces to the usual one, leading to a theory that can reproduce the $\Lambda$CDM model with the presence of a cosmological constant, as in Teleparallel gravity or GR. 

In addition, some particular cosmological solutions have been considered within the framework of $f(T)$ gravity with the presence of a conformally-invariant scalar field. For some particular scalar potentials, which remains the scalar field equation conformally invariant, power-law solutions and de Sitter solutions have been reconstructed. Note that for the particular case of a null scalar potential, a time dependent scalar field is capable of reproducing a pure de Sitter solution, which leads to a decreasing evolution of the scalar field, and gives rise to several de Sitter solutions (depending on the $f(T)$ action) that could be identified with the inflationary epoch and the dark energy phase. Furthermore, the $\Lambda$CDM model may be reproduced in standard teleparallel gravity $f(T)=T$ or in $\mathcal{C}$-gravity by the presence of the conformal scalar field, which would not contribute during the matter/radiation dominated epochs but would lead to an effective cosmological constant when the late-time acceleration starts, and the Hubble parameter is approximately constant. 

Hence, we have established some conformal properties within the framework of extended teleparallel theories, where some conformally-invariant models have been found, and the FLRW spacetimes have been studied. 

It would be very interesting to study the quantum properties of 
conformally-invariant theories in teleparallel gravity (for general review,
see~\cite{Buchbinder:1992rb}). 
In particular, we expect that as in curved spacetime with 
antisymmetric torsion~\cite{Buchbinder:1992rb}, 
the asymptotically conformal invariance may 
occur also in the conformal self-interacting scalar theory of Sec.~IV. 
Furthermore, the effect of conformal anomaly due to torsion on the FLRW 
cosmology should be studied in detail. It is quite possible that like in 
the usual GR with conformal matter, such conformal anomaly may 
induce the so-called anomaly-driven inflation.

\section*{Acknowledgments}

We are grateful to T.~Dereli for valuable comments and participation at the 
early stage of this work. 
Moreover, we would like to sincerely acknowledge the KMI visitor program at 
Kobayashi-Maskawa Institute for the Origin of Particles and the Universe (KMI), Nagoya University (S.D.O. and D.S-G), 
where we have completed this work. 
Furthermore, we are also grateful to Professor Shin'ichi Nojiri for 
very kind hospitality very much. 
In addition, D.S-G thanks Professor Rong-Jia Yang for useful discussions. 
The work is supported in part by 
the JSPS Grant-in-Aid for Young Scientists (B) \# 25800136 (K.B.) 
and 
MINECO (Spain), FIS2010-15640, 
AGAUR (Generalitat de Ca\-ta\-lu\-nya), contract 2009SGR-345, 
and MES project 2.1839.2011 (Russia) (S.D.O.). 
D.S-G also acknowledges the support from the University of the Basque Country and the URC financial support from the University of Cape Town (South Africa).

\end{document}